\documentclass[aps,prl,twocolumn,groupedaddress,showpacs,floatfix]{revtex4}
\usepackage{graphicx}
\usepackage{dcolumn}
\usepackage{bm}
\usepackage{amsfonts}
\usepackage{amsmath}

\begin{document}
\title{Emergence of atomic-density waves in a trapped Luther-Emery fermion gas}
\author{Gao Xianlong}
\author{Marco Polini}
\author{Mario P. Tosi}
\affiliation{NEST-INFM and Scuola Normale Superiore, I-56126 Pisa, Italy}
\author{Vivaldo L. Campo, Jr.}
\author{Klaus Capelle}
\affiliation{Departamento de F\'{\i}sica e Inform\'atica,
Instituto de F\'{\i}sica de S\~ao Carlos,
Universidade de S\~ao Paulo,
Caixa Postal 369, 13560-970 S\~ao Carlos, S\~ao Paulo, Brazil}
\date{\today}
\begin{abstract}
We present a novel and comprehensive microscopic study of Luther-Emery-paired phases in a 
strongly interacting atomic Fermi gas inside a parabolic trap and a one-dimensional ($1D$) optical lattice. 
Our work is based on a lattice version of density-functional theory, which uses as reference system the $1D$ homogeneous Hubbard model. We test our approach for repulsive interactions against Quantum Monte Carlo data and show that, for sufficiently strong attractions, an atomic-density wave (ADW) in the central portion of the trap breaks the discrete translational symmetry of the underlying lattice. We demonstrate that the emergence of an ADW has a dramatic impact on experimental observables such as the Fraunhofer diffraction pattern and the momentum distribution.
\end{abstract}
\maketitle

Strongly-interacting $1D$ quantum liquids are nowadays available in a variety of laboratory systems 
ranging from carbon nanotubes~\cite{carbon_nanotubes} to semiconductor nanowires~\cite{semiconducting_nanowires}, conducting molecules~\cite{Nitzan_03}, and gases in optical lattices~\cite{cold_atoms_low_D}. 
Chiral Luttinger liquids at fractional quantum Hall edges~\cite{chiral_ll} also provide a beautiful example of a $1D$ quantum liquid. The effective low-energy description of all these $1D$ systems is based on a harmonic theory of long-wavelength fluctuations~\cite{haldane} due to the interplay between topology and interactions. In the most interesting experimental situations translational invariance is broken by inhomogeneities of various types, 
such as Hall bar constrictions in the case of quantum Hall edges or trapping for ultracold atomic gases~\cite{bec_recent}. These strong external perturbations induce 
new length scales causing novel physical behaviors relative to the corresponding unperturbed model system. 

A powerful theoretical tool to study the interplay between interactions and inhomogeneous external fields of arbitrary shape is provided by density-functional theory (DFT)~\cite{dft,Giuliani_and_Vignale}, which is based on the Hohenberg-Kohn theorem and on the Kohn-Sham mapping onto an auxiliary noninteracting system. Many-body effects enter DFT {\it via} the exchange-correlation (xc) functional, which is often treated by the local-density approximation (LDA)~\cite{dft,Giuliani_and_Vignale}. The essence of LDA is to locally approximate the xc energy of the inhomogeneous system under study with that of an interacting homogeneous reference fluid, whose correlations are transferred by the LDA to the inhomogenous system. Most of the applications of DFT to inhomogeneous electronic systems use the interacting homogeneous electron gas (HEG)~\cite{Giuliani_and_Vignale} as the underlying reference fluid. The exact HEG xc energy is not known but can be calculated to a high degree of numerical precision with the help of Quantum Monte Carlo (QMC) data. However, there are a few interesting examples in the literature~\cite{general} in which either the reference system is not the HEG or the auxiliary system of the Kohn-Sham mapping is not an assembly of noninteracting particles. In particular, inhomogeneous $1D$ fermionic systems appear as an example in which it is appropriate to change the reference system to one that possesses ground-state (GS) Luttinger-liquid rather than Fermi-liquid-type correlations (see Ref.~\onlinecite{lima_2003} and references therein to earlier work). 

In this Letter we present the first such DFT study of trapped ultracold Fermi gases. We focus our attention on a two-component Fermi gas in a $1D$ optical lattice with either repulsive or attractive inter-component interactions and in the presence of a static external potential. Creating these types of systems experimentally seems to be within the reach of present-day technology~\cite{experiments_cold_fermi_gases}. Studies of this model have only been carried out in the case of repulsive interactions~\cite{rigol,machida_2004,xia_ji_2005} whereas the main focus of this Letter is on attractive interactions. We use a lattice version of DFT within an LDA based on the $1D$ homogeneous Hubbard model (HHM)~\cite{lieb_wu}, which transfers Luttinger-Mott and Luther-Emery-type~\cite{luther_emery} correlations to the inhomogeneous gas. A novel type of ground state is found at strong attractive coupling.

We consider a Fermi gas with $N_{\rm f}$ particles hopping on a $1D$ lattice with $N_{\rm s}$ lattice sites.
Each site is labeled by the discrete coordinate $\{z_i=i, i\in[1,N_{\rm s}]\}$. We assume that the system can be described by a single-band Hubbard Hamiltonian~\cite{jaksch_98},
\begin{eqnarray}\label{eq:hubbard}
{\hat {\cal H}}&=&-\sum_{i,j}\sum_\sigma
t_{i,j}\left[{\hat c}^{\dagger}_{\sigma}(z_i){\hat c}_{\sigma}(z_j)+{\rm H}.{\rm c}.\right]\nonumber\\
&+&U\sum_{i}\,{\hat n}_{\uparrow}(z_i){\hat n}_{\downarrow}(z_i)
+\sum_{i}V_{\rm ext}(z_i){\hat n}(z_i)\,.
\end{eqnarray}
Here $t_{i,j}=t>0$ if $i,j$ are nearest-neighbor sites and zero otherwise, $\sigma=\uparrow,\downarrow$ is 
a pseudospin-$1/2$ degree of freedom (hyperfine-state label), ${\hat n}_\sigma(z_i)= {\hat c}^{\dagger}_{\sigma}(z_i){\hat c}_{\sigma}(z_i)$ is the pseudospin-resolved site occupation operator normalized to the number of particles with pseudospin $\sigma$, $\langle \sum_i {\hat n}_\sigma(z_i)\rangle=N_\sigma$, and ${\hat n}(z_i)=\sum_\sigma {\hat n}_\sigma(z_i)$ is the total site occupation with $\langle \sum_i{\hat n}(z_i)\rangle=N_{\rm f}$. Finally $V_{\rm ext}(z)$ is an external potential that we take later below as representing a harmonic trap.

If $V_{\rm ext}(z)=0$ the Hamiltonian in Eq.~(\ref{eq:hubbard}) reduces to a $1D$ HHM 
that can be exactly solved using the Bethe-{\it Ansatz} (BA) technique for both repulsive ($U>0$) and attractive ($U<0$) interactions~\cite{lieb_wu}. The properties of the system are determined by three parameters: (i) the filling factor $n=N_{\rm f}/N_{\rm s}$, (ii) the pseudospin polarization $s=(N_\uparrow-N_\downarrow)/(2N_{\rm s})$, and (iii) the dimensionless coupling constant $u=U/t$. The energy $\varepsilon(n,s;u)$ per site can be calculated by solving a set of nonlinear coupled integral equations. In agreement with the Lieb-Mattis theorem~\cite{lieb_mattis} the GS has $s=0$. In what follows the GS energy at $s=0$ will be denoted by $\varepsilon_{\rm \scriptscriptstyle GS}(n,u)=\varepsilon(n,0;u)$. For $u>0$ the $1D$ HHM describes a Luttinger liquid~\cite{schulz_1990} if $n\neq 1,2$: the two metallic GS branches for $n<1$ and $n>1$ are connected by particle-hole symmetry, 
$
\varepsilon_{\rm \scriptscriptstyle GS}(n,u)=(n-1)U+\varepsilon_{\rm \scriptscriptstyle GS}(2-n,u)
$. 
At half filling, {\it i.e.} for $n=1$, the GS is a Mott insulator for every $u\neq 0$, as signaled by a cusp in $\varepsilon_{\rm \scriptscriptstyle GS}$ and by a gap in the charge excitation spectrum, 
while for $n=2$ the GS is a band insulator. 
Turning to the case $u<0$, the $1D$ HHM describes a Luther-Emery liquid~\cite{luther_emery}, which exhibits a gap $\Delta_{\rm S}(n)$ in the spin excitation spectrum and no charge gap. Due to an exact symmetry of the $1D$ HHM, the GS energy for $u<0$ is given by the energy of a repulsive $1D$ HHM at half filling and variable spin polarization $s(n)=(1-n)/2$,
$
\varepsilon_{\rm \scriptscriptstyle GS}(n,-|u|)=-|U|\,n/2+\varepsilon(1,s(n);|u|)
$.
The Bardeen-Cooper-Schrieffer (BCS) approximation 
has been shown~\cite{marsiglio_1997} to provide a good description of the GS energy but a rather inaccurate 
and even qualitatively wrong description of the gap $\Delta_{\rm S}(n)$. 

In the presence of a confining external potential ${\hat {\cal H}}$ cannot be diagonalized exactly. 
We calculate the GS properties of ${\hat {\cal H}}$ in Eq.~(\ref{eq:hubbard}) by a lattice version of DFT, the so-called site-occupation-functional theory (SOFT). SOFT was introduced in a pioneering paper by Gunnarsson and Sch\"onhammer~\cite{soft} in order to study the so-called band-gap problem~\cite{Giuliani_and_Vignale} in the context of {\it ab initio} theories of fundamental energy gaps in semiconductors and insulators. Within SOFT the exact GS site occupation $n_{\rm \scriptscriptstyle GS}(z_i)=\langle{\rm GS}|{\hat n}(z_i)|{\rm GS}\rangle$ can be obtained by solving self-consistently the Kohn-Sham Schr\"odinger equations
\begin{equation}\label{eq:sks}
\sum_{j}[-t_{i,j}+v_{\rm \scriptscriptstyle KS}(z_i;[n_{\rm \scriptscriptstyle GS}(z_i)])\delta_{ij}]\varphi_\alpha(z_j)=\varepsilon_\alpha\varphi_\alpha(z_i)\,,
\end{equation}
with $v_{\rm \scriptscriptstyle KS}(z_i;[n_{\rm \scriptscriptstyle GS}(z_i)])=U n_{\rm \scriptscriptstyle GS}(z_i)/2+v_{\rm xc}(z_i)+V_{\rm ext}(z_i)$, together with the closure
\begin{equation}\label{eq:closure}
n_{\rm \scriptscriptstyle GS}(z_i)=\sum_{\alpha, {\rm occ.}}\Gamma_\alpha\left|\varphi_\alpha(z_i)\right|^2\,.
\end{equation}
Here the sum runs over the occupied orbitals and the degeneracy factors $\Gamma_\alpha$ satisfy the sum rule 
$\sum_\alpha \Gamma_\alpha=N_{\rm f}$. The first term in the effective Kohn-Sham potential is the Hartree potential, while $v_{\rm xc}(z_i;[n_{\rm \scriptscriptstyle GS}(z_i)])=\delta E_{\rm xc}[n(z_i)]/\delta n(z_i)|_{\rm \scriptscriptstyle GS}$ is the xc potential defined by the functional derivative of the xc energy $E_{\rm xc}[n(z_i)]$ evaluated at the GS site occupation. The total GS energy of the inhomogeneous system is given by $E_{\rm \scriptscriptstyle GS}[n_{\rm \scriptscriptstyle GS}(z_i)]=\sum_\alpha\Gamma_\alpha\varepsilon_\alpha-\sum_i v_{\rm xc}(z_i)n_{\rm \scriptscriptstyle GS}(z_i)-\sum_i U n^2_{\rm \scriptscriptstyle GS}(z_i)/4+E_{\rm xc}[n_{\rm \scriptscriptstyle GS}(z_i)]$.

Equations~(\ref{eq:sks}) and (\ref{eq:closure}) provide a formally exact scheme to calculate $n_{\rm \scriptscriptstyle GS}(z_i)$ and $E_{\rm \scriptscriptstyle GS}[n_{\rm \scriptscriptstyle GS}(z_i)]$, but $E_{\rm xc}$ and $v_{\rm xc}$ need to be approximated. The LDA has been shown to provide an excellent description of the GS properties of a variety of inhomogeneous systems~\cite{dft}. In the following we employ a BA-based LDA (BALDA) functional~\cite{lima_2003},
\begin{equation}\label{eq:balda}
v^{\rm \scriptscriptstyle BALDA}_{\rm xc}(z_i;[n_{\rm \scriptscriptstyle GS}(z_i)])=\left.v^{\rm hom}_{\rm xc}(n,u)\right|_{n\rightarrow n_{\rm \scriptscriptstyle GS}(z_i)}\,.
\end{equation}
Here the xc potential of the $1D$ HHM is defined by $v^{\rm hom}_{\rm xc}(n,u)=
\partial_n\left[\varepsilon_{\rm \scriptscriptstyle GS}(n,u)-k(n)\right]-Un/2$, with 
$k(n)=-4t\sin{(n\pi/2)}/\pi$ being the kinetic energy of the noninteracting gas. We should point out two appealing features of Eq.~(\ref{eq:balda}) from the formal DFT viewpoint. First of all, the xc potential of the reference $1D$ HHM can be exactly calculated from BA integral equations~\cite{footnote1}. Secondly, the Mott cusp in $\varepsilon_{\rm \scriptscriptstyle GS}(n,u)$ for repulsive interactions is responsible for a discontinuity $\Delta_{\rm xc}(U)$ in $v^{\rm hom}_{\rm xc}$ at $n=1$, $\Delta_{\rm xc}(U)=U-2\lim_{n\rightarrow 1^-} \partial_n \varepsilon_{\rm \scriptscriptstyle GS}(n<1,u)$. As a consequence and contrary to the HEG-based LDA, the xc potential in Eq.~(\ref{eq:balda}) naturally possesses a derivative discontinuity~\cite{Giuliani_and_Vignale,lima_2002}. 

\begin{figure}
\includegraphics[width=1.00\linewidth]{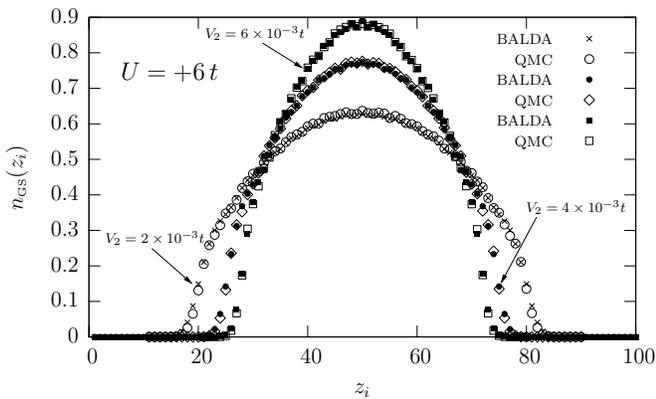}
\caption{Ground-state site occupation $n_{\rm \scriptscriptstyle GS}(z_i)$ as a function of site $z_i$ 
for a paramagnetic repulsive Fermi gas with $u=+6$ and in the presence of a harmonic potential $V_{\rm ext}(z_i)=V_2(z_i-N_{\rm s}/2)^2$, with $V_2/t=2\times 10^{-3}, 4\times 10^{-3}$, and $6\times 10^{-3}$. In this Figure and in all the others we have chosen $N_{\rm f}=30$ and $N_{\rm s}=100$. The results of the BALDA self-consistent scheme are compared with the QMC data of Ref.~\onlinecite{rigol}.\label{fig:one}}
\end{figure}

\begin{figure}
\includegraphics[width=1.00\linewidth]{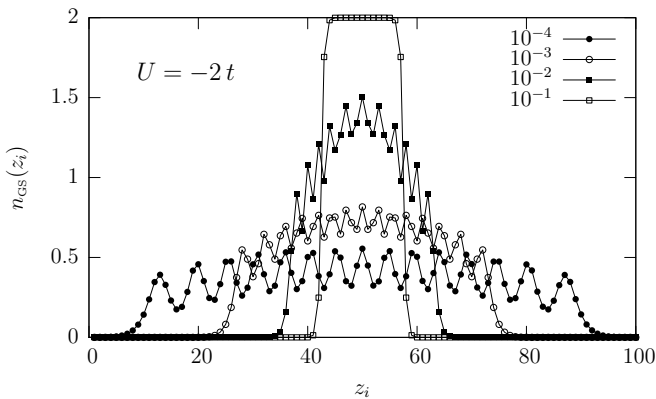}
\caption{Ground-state site occupation $n_{\rm \scriptscriptstyle GS}(z_i)$ as a function of site $z_i$ 
for a paramagnetic attractive Fermi gas with $u=-2$ and in the presence of a harmonic potential, with $V_2/t$ increasing from $10^{-4}$ to $10^{-1}$. The thin solid lines are just a guide for the eye. In the bulk region ($40\leq z_i\leq 60$) and for $V_2/t=10^{-4}$, $n_{\rm \scriptscriptstyle GS}(z_i)$ can be well fitted to an ADW of the form $n_{\rm \scriptscriptstyle GS}(z_i)=n_0+n_1\cos{(2\pi z_i/\lambda_{\rm \scriptscriptstyle ADW}+\varphi)}$ with $n_0=0.43$, $n_1=0.12$, $\lambda_{\rm \scriptscriptstyle ADW}=4.70$ (to be compared with the prediction $\lambda_{\rm \scriptscriptstyle ADW}=2/n=6.67$ for the homogeneous gas~\cite{luther_emery}), and $\varphi=2.17$.\label{fig:two}}
\end{figure}

\begin{figure}
\includegraphics[width=1.00\linewidth]{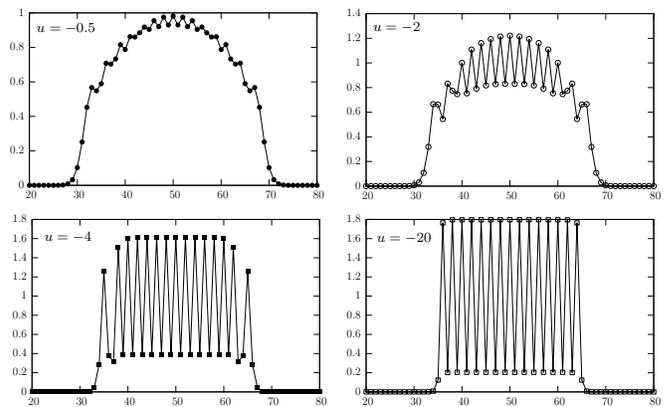}
\caption{Ground-state site occupation $n_{\rm \scriptscriptstyle GS}(z_i)$ as a function of site $z_i$ for a paramagnetic attractive Fermi gas in the presence of a harmonic external potential with $V_2/t=4 \times 10^{-3}$ and for $u=-0.5,-2,-4$, and $-20$. The thin solid lines are just a guide for the eye.\label{fig:three}}
\end{figure}

We demonstrate the accuracy of the BALDA scheme for trapped $1D$ Fermi gases in Eqs.~(\ref{eq:sks})-(\ref{eq:balda}) by showing in Fig.~\ref{fig:one} some illustrative results for the GS site occupation of harmonically trapped repulsive fermions against the state-of-the-art QMC data of Rigol {\it et al.}~\cite{rigol}. 
The agreement between our theory and the QMC results is clearly excellent~\cite{footnote2}. We now turn to describe how the density profile changes when $U$ becomes negative. In Figs.~\ref{fig:two} and~\ref{fig:three} we report our theoretical predictions for the density profiles of attractive fermions. We clearly see from these figures that the consequences of Luther-Emery pairing in a confined Fermi gas are dramatic. The GS site occupation is a finite-size density wave, {\it i.e.} an oscillating function of $z_i$ that reflects the tendency of atoms with antiparallel pseudospins to form spin-singlet dimers. 

In Fig.~\ref{fig:two} we illustrate the role of the external harmonic potential.
For a very weak confinement the central part of the trap is occupied by a small-amplitude ADW, but in a sufficiently strong trap a region of doubly-occupied sites is formed at the center of the trap. In Fig.~\ref{fig:three} we show the crossover from the weak-coupling BCS regime to a strong-coupling regime where the spin-singlet dimers are tightly bound.
On increasing the magnitude of the atom-atom attraction the amplitude of the Luther-Emery oscillations increases dramatically, giving rise to a large-amplitude ADW with period $\lambda_{\rm \scriptscriptstyle ADW}\approx 2$. The tendency of the system to form stable dimers is also confirmed numerically by a calculation of the pair-binding energy ${\cal E}_{\rm \scriptscriptstyle P}=E_{\rm \scriptscriptstyle GS}(N_{\rm f}+2)+E_{\rm \scriptscriptstyle GS}(N_{\rm f})-2E_{\rm \scriptscriptstyle GS}(N_{\rm f}+1)$, which turns out to be negative (see Table~\ref{table:one}). 
\begin{table}
\caption{Ground-state and pair-binding energies for a paramagnetic attractive Fermi gas in a harmonic external potential for $u=-0.5,-2,-4$, and $-20$. Note that we are able to handle also large systems with our SOFT scheme.\label{table:one}}
\begin{ruledtabular}
\begin{tabular}{cccccc}
$u$ & $V_2/t$ & $N_{\rm s}$ & $N_{\rm f}$ & $E_{\rm \scriptscriptstyle GS}/(tN_{\rm s})$& ${\cal E}_{\rm \scriptscriptstyle P}/t$\\
$-0.5$  & $4\times 10^{-3}$  & $100$  &$30$  &$-0.358$ &$-0.009$\\
$$      & $10^{-4}$  				 & $300$  &$90$  &$-0.489$ &$-0.003$\\ \hline
$-2$    & $4\times 10^{-3}$  & $100$  &$30$  &$-0.477$ &$-0.092$\\
$$      & $10^{-4}$          & $300$  &$90$  &$-0.583$ &$-0.039$\\ \hline
$-4$    & $4\times 10^{-3}$  & $100$  &$30$  &$-0.707$ &$-0.861$\\
$$      & $10^{-4}$          & $300$  &$90$  &$-0.786$ &$-0.791$\\ \hline
$-20$   & $4\times 10^{-3}$  & $100$  &$30$  &$-3.053$ &$-1.816$\\
$$      & $10^{-4}$          & $300$  &$90$  &$-3.120$ &$-1.509$\\
\end{tabular}
\end{ruledtabular}
\end{table}

\begin{figure}
\includegraphics[width=1.00\linewidth]{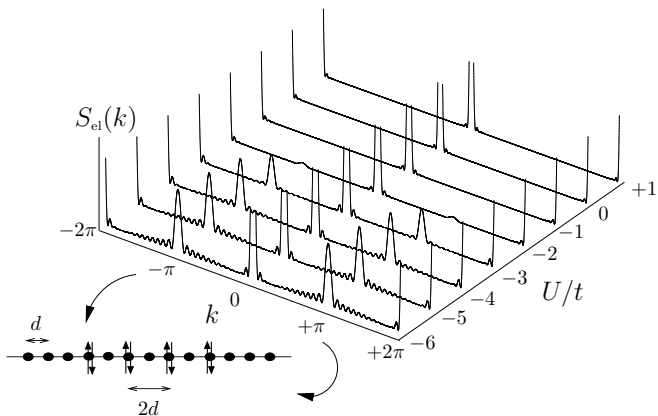}
\caption{The Fraunhofer structure factor $S_{\rm el}(k)$ as a function of wavenumber $k$ in the range $-2\pi\leq k \leq 2\pi$, for varying $u$ in a paramagnetic Fermi gas inside a harmonic trap with $V_2/t=4 \times 10^{-3}$. For clarity we have limited the $z$ values to the interval $[0,0.3]$. 
The satellite peak at $k=\pm\pi$ has magnitude $0.014$ for $u=-2$ and 
$0.28$ for $u=-6$. The inset shows a pictorial description of the unit-cell doubling, which drives the emergence of the satellites.\label{fig:four}}
\end{figure}

We proceed to illustrate the impact of the above ADWs on some measurable quantities. The key point is that the real-space Luther-Emery ADWs with their intrinsic periodicity $\lambda_{\rm \scriptscriptstyle ADW}\neq 1$ break the lattice symmetry and transfer structure to momentum space at wavenumbers $k\approx k_{\rm \scriptscriptstyle ADW}=\pm 2\pi/\lambda_{\rm \scriptscriptstyle ADW}$, away from the reciprocal lattice vectors $G_\nu=2\pi\nu$, $\nu=0,\pm1, \pm 2,...$. As an illustrative example we show in Fig.~\ref{fig:four} the elastic contribution to the light-scattering diffraction pattern~\cite{vignolo_2001}, {\it i.e.} the Fraunhofer structure factor $S_{\rm el}(k)=N^{-2}_{\rm f}\left|\sum_{i}\exp{(-ikz_i )}n_{\rm \scriptscriptstyle GS}(z_i)\right|^2$, in the case of a relatively strong harmonic confinement [note that $S_{\rm el}(G_\nu)=1$ and $S_{\rm el}(k+G_\nu)=S_{\rm el}(k)$]. In this case, bulk ADWs with wavenumber $k_{\rm \scriptscriptstyle ADW}=\pm\pi$ characterize the GS density profile for a sufficiently strong attraction. The emergence of a satellite peak at $k=\pm \pi$ for $u\lesssim -2$ is evident in Fig.~\ref{fig:four}. 

For the same parameters employed in Fig.~\ref{fig:four} the momentum distribution function of an attractive Fermi gas possesses a satellite peak at $k=\pm\pi$, which could be observed in a time-of-flight experiment. 
In addition to light scattering and time-of-flight-type experiments, trapping in a ring-shaped optical lattice~\cite{amico_2005} can also unambiguously reveal the Luther-Emery phases. The response of the system to an Aharonov-Bohm flux~\cite{twisted_boundary_conditions} can be used to characterize microscopically the nature of the ground state. In fact, if the GS has a spin gap as in the Luther-Emery phase, it has been shown~\cite{seidel_2005} that for a large system the period of the GS energy as a function of the Aharonov-Bohm flux is $\pi$, exactly half the expected one. Thus ultracold Fermi gases with attractive interactions could be a highly tunable system in which it is possible to observe direct evidence of Luther-Emery pairing. 

In conclusion we remark that our DFT scheme is also applicable to a number of other experimentally interesting systems such as non-paramagnetic gases, gases under the effect of time-dependent external potentials, and $1D$ quantum electron liquids.  

\begin{acknowledgments}
This work was partially supported by MIUR, FAPESP, and CNPq.
We thank Marcos Rigol for providing us with his QMC  
results and the ``HLR-Stuttgart (Project DynMet)" for the allocation of 
computer time. We also thank Dr. P. Vignolo for very useful discussions.
\end{acknowledgments}


\begin{thebibliography}{99}
\bibitem{footnote2}
	The agreement worsens when local insulating regions appear within the metallic phase, 
	since it becomes difficult to obtain a self-consistent solution of the Kohn-Sham equations 
	for the density plateaus observed in Ref.~\onlinecite{rigol}. 
	In the present work, however, we are concerned only with the metallic 
	and Luther-Emery phases, where convergence is not a problem.
\end{thebibliography}
\end{document}